\documentclass{book}

\usepackage[yearbook]{ivc} % in some cases the option is not "yearbook" but "grauereihe"

\begin{document}

\markboth{Dennis Dieks}{Uniqueness of entropy: micro vs.\ macro}

\autorenaufsatz{Dennis Dieks}

\titelaufsatz{Is There a Unique Physical Entropy? Micro versus Macro}

\sectioneins{Entropy in Statistical Physics}

The concept of entropy has become common even in everyday language, in which it rather vaguely refers to disorder, loss of ``energy'', waste and dissipation. Users of the concept generally take it for granted that in the background there is a precise scientific notion, with which one should be able to justify, at least in principle, this informal parlance. It is therefore perhaps surprising to find that even in the exact sciences entropy is a multi-faceted concept. It is perhaps least controversial in probability and information theory, at least as far as its mathematical expression is concerned: $S = -\sum_i p_i \ln p_i$ is the generally accepted formula for the entropy $S$ of a probability distribution $\{p_i\}$. But even in the mathematical fields of probability and information theory the exact significance of entropy, and the role that it can play in, e.g., decision theoretical contexts, remains to some extent controversial. One might hope that this will be different once the use of entropy in physics is considered. After all, in physics one expects that the term ``entropy'' will correspond to something that is accessible to measurement--and drastic differences of opinion about something that can be measured would be surprising. It is this physical entropy, in statistical physics and in thermodynamics, that we shall be concerned with in this paper.

For the case of $M$ equiprobable events, $p_i = 1/M$, the formula $S = -\sum_i p_i \ln p_i$ reduces to $S = \ln M$. Essentially, this is the famous formula $ S = k \ln W$ that can be traced back to Ludwig Boltzmann's seminal 1877 paper about the relation between the second law of thermodynamics and probability theory (Boltzmann 1877, 2001). The constant $k$ (Boltzmann's constant) is merely introduced in order to fix the unit; and $W$ is the number of microstates corresponding to a given macrostate---it is a number of possibilities like $M$ in the earlier formula. The macrostate is defined by macroscopic quantities like pressure, volume and temperature (in the case of a gas in a container); $W$ is the number of microscopic states, characterized by the positions and velocities of the atoms or molecules in the gas, that each give rise to the same values of these macroscopic quantities and in this sense belong to the same macrostate. Boltzmann's entropy thus is basically the earlier introduced $S$ for the case of a probability distribution that assigns equal probabilities to all microstates belonging to a given macrostate. Since the microstates can be represented as points in the phase space of the physical system, the formula $ S = k \ln W$ tells us that the entropy of a macrostate is proportional to the logarithm of the volume in microscopic phase space that corresponds to the macrostate.

A paradigmatic and simple application of $ S = k \ln W$ is the case of $N$ classical particles (atoms or molecules), each of which can be in any one of $X$ possible states. In this case we find $W = X^N$, and therefore $S = kN\ln X$.

\sectioneins{Entropy in Thermodynamics}

In thermodynamics, physical systems are considered from a purely macroscopic point of view. In the case of a gas in a container one looks at changes in macroscopically measurable quantities when the pressure $P$, volume $V$ and temperature $T$ are made to vary. An essential result, at the basis of the so-called second law of thermodynamics, is that different ways of going from one macroscopic state $A$ to another macroscopic state $B$ (for example, by either first compressing and then cooling, or doing these things in reversed order) are generally associated with different amounts of exchanged heat $\Delta Q$. The heat content of a physical system is therefore not a quantity fixed by its macroscopic state: it is not a \textit{state function}. However, the quantity $\int_{A}^{B} dQ/T$, i.e.\ the exchanged heat divided by the temperature, integrated along a path from $A$ to $B$ (in the macroscopic state space) that represents a \textit{reversible} process, \emph{is} path-independent. That means that $\int_O dQ/T$ does define a state function (the choice of the fiducial state $O$ defines the zero of this function; different choices of $O$ lead to functions that differ by a constant). It is this macroscopic state function that defines the thermodynamic entropy: $S \equiv \int dQ/T$.

Boltzmann's seminal 1877 idea was that the statistical entropy $ S = k \ln W$ (Boltzmann himself used another notation) is the microscopic counterpart of the thermodynamic entropy. Each macroscopic state corresponds to a volume in phase space on the micro level, namely the volume occupied by all those microstates that give rise to the macrostate in question; and the logarithm of this volume represents (apart from immaterial constants) the thermodynamic entropy of the macrostate.

\sectioneins{A discrepancy}%numbers may be omitted

If the micro and macro entropies stand for one and the same physical quantity, the two entropies should obviously depend in exactly the same way on all variables. As it turns out, however, this necessary requirement is not fulfilled. The macro-entropy is \emph{extensive}: if we scale up a physical system by increasing its particle number, its energy and its volume by a factor $\lambda$, its entropy will increase by this same factor $\lambda$. In other terms, $S(\lambda N,\lambda V,\lambda E) = \lambda S(N,V,E)$. But the micro-entropy as defined above is not extensive.

To see this, imagine two gas-filled chambers of the same volume, separated by a
partition. Both chambers contain equal amounts of the same gas in
equilibrium, consisting of the same number $N$ of particles. Both parts have the same total energy,
temperature $T$ and pressure. Now the partition is removed. What happens to the entropy?

According to thermodynamics the entropy remains the same, because
the macroscopic properties of the gases do not
change. Smooth removal of the partition is a reversible process without heat transfer; therefore $S_{A}=S_{B}$, with $A$ and $B$ the macrostates before and after the removal, respectively. So the total entropy of the double amount of gas, without the partition, is the same as the combined entropy of the two original volumes, i.e. double the entropy of each of the two halves (in this it has been taken for granted that the entropy of several isolated systems is additive---see van Kampen 1984).

However, from the microscopic point of view, the number of available
states per particle doubles when the partition is taken out: each particle now has twice as
much phase space available to it as it had before. If the number of available states per particle was $X$ with the partition still in place, it becomes $2X$ after the removal of the partition. This means that the
number of microstates goes up, from $W_A=X^{2N}$ to $W_B=(2X)^{2N}$, which
corresponds to an entropy difference $S_{B}-S_{A} = 2kN \ln2$.

This discrepancy, known as (a version of) the Gibbs paradox, shows that although the thermodynamic entropy is extensive (it doubles when the amount of gas is doubled), the statistical mechanical entropy is not. If we think that there is one and only one physical entropy, this difference between the two approaches signals a problem that needs to be solved. Only one of the two expressions can be right in this case, and since we can directly measure the thermodynamic entropy, and verify its value, it seems clear that the Boltzmann formula $S=k \ln W$ must be wrong. There are two approaches in the literature that take this line. Both claim that fundamental reasoning, starting from first principles on the microscopic level, will not lead to the expression $S=k \ln W$, but instead to the formula $S = k \ln W/N!$, with $N$ the number of particles. This modification of the expression is sufficient to remove our discrepancy.

Remarkably, the two approaches have diametrically opposite starting points: the first, traditional one claims that the \emph{indistinguishability} of particles of the same kind must be taken into account and that this necessitates the insertion of $1/N!$. The second approach says that the \emph{distinguishability} of classical particles has been neglected.

\sectioneins{The standard ``solution'': indistinguishability of particles of the same kind}

The traditional way of responding to the discrepancy between micro and macro entropy is to point out that the particles (atoms or molecules) in the two gas chambers are ``identical'': since they are all atoms or molecules of the same gas, they all possess the same intrinsic properties (charge, mass, etc.). Therefore, a permutation of two or more of these particles should not lead to a new state: it cannot make a difference whether particle $1$ is in state $a$ and particle $2$ in state $b$, or the other way around. Both cases equally represent one particle in $a$ and one particle of the same type in $b$. If we go along with this, the number of microstates $W$ must be adjusted: for a system of $N$ identical particles it must be a factor $N!$
smaller than what we
supposed above. When we now redo the calculation, the removal of the partition between the two chambers changes $W$ from $W_A=X^{2N}/(N!)^2$ to
$W_B=(2X)^{2N}/(2N)!$. With the help of Stirling's approximation for the factorial
it follows that, in the so-called thermodynamic limit $N \rightarrow
\infty$, $W_B=W_A$. So the total entropy does not change when the partition is taken out: the resulting double-volume amount of gas has double the entropy of each of the separate chambers. This removes the discrepancy between
statistical mechanics and thermodynamics.

According to several authors and textbooks, in the final analysis quantum theory is
needed for justifying this solution of the Gibbs paradox (see e.g.\
Schr\"{o}dinger 1948, Huang 1963, Wannier 1966, Sommerfeld 1977,
Schroeder 2000, Ben-Naim 2007). Indeed, classical particles are
always distinguishable by their positions, which are strictly correlated to their individual trajectories. These trajectories, in other words the particles' histories, individuate the particles: if we give the particles names on the basis of their positions at one instant, these names persist through time. So the situation in which particle $1$ is in state $a$ at a later time is different from the situation in which $2$ is in $a$. It is therefore not self-evident in classical statistical mechanics that we should divide by $N!$. Identical quantum particles, on the other
hand, seem indistinguishable in the required sense from the start,
because quantum states of systems of identical particles must
either be symmetrical under permutation (bosons) or
anti-symmetrical (fermions): exchange of particles leaves the
state therefore invariant (apart from a global phase factor) and the
multiplicity $N!$ never enters.

If this argument were correct, then the non-extensivity of the Boltzmann entropy would show
that classical physics is inconsistent and that the world must be quantum mechanical. But obviously, it is hard to
believe that simple considerations about doubling amounts of gases could produce such fundamental insights. Unsurprisingly
therefore, doubts have been expressed concerning the
just-mentioned traditional solution of the paradox. For example, some authors
have claimed that identical classical particles are also fully
indistinguishable, and that this justifies the factor $1/N!$ without any
recourse to quantum mechanics (e.g., Hestenes 1970, Fujita 1991,
Nagle 2004, Saunders 2006).

In the next section we shall take a closer look at whether the permutation of classical particles does or does not make a difference for the microstate.

\sectioneins{Permutations of ``identical'' classical particles}

We already observed that classical particles can be named and
distinguished by their different histories. A process in which two classical particles of the same kind are interchanged can therefore certainly produce a
different microstate. Indeed, imagine a situation in which there
is one particle at position $x_1$ and one particle at position
$x_2$, and in which at a later instant there is again one particle
at $x_1$ and one at $x_2$; suppose that their respective
momenta are the same as before. What has happened in the meantime?
There are two possibilities: either the particle that was first at
$x_1$ is later again at $x_1$ and the particle that was first at
$x_2$ is later again at $x_2$, or the particles have exchanged
their positions. The latter case would clearly be different from
the former one: it corresponds to a different physical process.
Although it is true that the two final situations cannot be distinguished on the
basis of their instantaneous properties, their different
histories show that the particle at $x_1$ in one final situation
is not the same as the particle at $x_1$ in the other final
situation.

These remarks seem trivial; so what is behind the denial by some authors that identical classical particles can be
distinguished and that permutations give rise to different
microstates? One reason is that there is an ambiguity in the
meaning of the terms ``distinguishable'' and ``permutation''.
Consider the following statements: ``Two particles are
distinguishable if they can always be selectively separated by a
filter'' (Hestenes 1970); ``Two particles are distinguishable
if they are first identified as 1 and 2, put into a small box,
shaken up, and when removed one can identify which particle was
the original number 1'' (Nagle 2004). With \emph{these}
definitions of distinguishability particles of the same kind
are indeed indistinguishable. The concept of
``permutation'' can be interpreted in a similar way. Consider again
the microstate of two particles of the same kind, one at $x_1$ and
another at $x_2$. If the particle at $x_2$ were at $x_1$ instead,
and the particle at $x_1$ were at $x_2$, with all properties
interchanged, there would be no physical differences, neither from
an observational point of view nor from the viewpoint of theory.
One can therefore certainly maintain that the two
situations are only two different descriptions (using different ways of assigning indices) for one and the same physical situation (Fujita 1991).

But this is a different kind of permutation from the physical
exchange we considered before. In our first example the particles
\emph{moved} from $x_1$ to $x_2$ and \textit{vice versa}.
Trajectories in space-time connected the initial state to the
permuted state. By contrast, in the alternative reading of
``permutation'' just mentioned, the exchange is not a physical
process at all. Instead, it is an instantaneous swapping that
occurs in our thought; it exchanges nothing but indices and does
not need trajectories.

A similar sense of ``permutation'' is employed
by Saunders (Saunders 2006). Consider one particle $a$ that follows trajectory
1 and another particle $b$ of the same kind that follows trajectory 2. Now imagine the case in which particle $a$
followed trajectory 2 and particle $b$ followed trajectory 1. This exchange would not make any difference for the physical situation.
As before, the states before and after a permutation of this kind are not connected by
a physical process. A permutation in this sense swaps a supposedly existing abstract ``identity'' (formally
represented by the particle indices ``1'' and ``2'', respectively)
that is completely independent of the physical characteristics of the
situation.

The upshot of these considerations is that if ``permutation'' is
understood as a physical exchange in which trajectories in
space-time connect the initial state to the permuted state, then
permutations give rise to physically different possibilities, in the
sense of different physical processes.
If ``permutation'' is however understood in a different way, then
it may well be true that such permutations are not
associated with any physical differences and so do not lead to a new microstate.

Let us now consider which kind of
permutations is relevant to statistical mechanics---physical
exchanges, with connecting trajectories, or swapping indices? Which kind of permutations determines the number of
microstates $W$?

Remember our two gas-filled chambers, each containing $N$
identical particles. Before the removal of the partition the number of
available states per particle is $X$. After the partition has been
removed, the number of available states has become $2X$. The reason is
that after the partition's removal it has become possible for the
particles to \emph{move} to the other chamber. The doubling of the
number of available microstates thus expresses a physical freedom
that did not exist before the partition was taken away: trajectories have become possible from the
particles' initial states to states in the other chamber.

In contrast, even with the partition in place we could consider,
in thought, the permutation of ``particle identities'', or indices, from the left and right
sides, respectively---but such permutations are never taken into account in the
calculation of the number of microstates. Nor do we consider permutations
with particles of the same kind outside of the container, obviously. In
other words, the relevant kind of permutations are physical
exchanges, not the abstract swapping of indices or identities.

To completely justify the answer that accessibility via a real
physical process is the determining
factor in the calculation of the number of microstates, we would have to
go deeper into the foundations of statistical mechanics. Here, we only
mention that one important approach in this area is the ergodic theory, in which the probability of a macrostate is
argued to be proportional to the associated volume in phase space on the grounds that
this volume is proportional to the amount of time a system will actually
dwell in that part of phase space that corresponds to the
macrostate in question. Clearly, this idea only makes sense if the
microstates in this part of the phase space are actually
accessible via physical trajectories: microstates that give rise to
the same macrostate but cannot be reached from the initial
situation through the evolution of the system are irrelevant for
the macrostate's probability---they do not play a role at all.

It is true that the original form of the ergodic hypothesis
(according to which all microstates are actually visited in a
relatively short time) has proven to be untenable, but this does
not impugn the basic idea that accessibility is the criterion for
the relevance of microstates. The multiplicities that occur in
more modern and more sophisticated approaches to the foundations
of statistical mechanics are the same as those of the original
ergodic theory.

We can therefore conclude that in classical statistical mechanics
the relevant number of microstates is sensitive to the number of ways
this macrostate can be reached via physical processes, i.e. different paths in phase space.
Given $N$ particles, there are generally $N!$ different ways in which the particles that have been numbered at some initial time can be distributed in a state at a later time. These permutations represent different physical possibilities, corresponding to different physical processes. Dividing by $N!$ is therefore unjustified when we calculate the numbers of microstates that can be realized by classical particles of the same kind\footnote{A more detailed discussion should also take into account that the division by $N!$ is without significance anyway as long as $N$ is constant: in this case the only effect of the division is that the entropy is changed by a constant term $\ln N!$, see (Versteegh 2011).}.

\sectioneins{An alternative ``solution'': distinguishability of particles of the same kind}

In a number of recent publications, Swendsen has proposed an alternative line of reasoning that leads to the entropy formula $S=k \ln W/N!$; he claims that this derivation, rather than the standard accounts, captures the essence of Boltzmann's 1877 ideas (e.g., Swendsen 2002, Swendsen 2008, Swendsen 2012). Swendsen's strategy is to calculate the entropy of a system by considering it as a part of a bigger, composite system; and then to look at the probabilities of microstates of this composite system. Boltzmann's 1877 definition is interpreted as saying that the logarithm of this probability distribution is the entropy of the composite system (apart from multiplicative and additive constants).

Let us illustrate Swendsen's approach by combining a system consisting of a gas of volume $V_1$ and particle number $N_1$ with a second gas of the same kind, with volume $V_2$ and particle number $N_2$. Let us denote the total volume by $V$: $V= V_1 + V_2$. The total number of particles, $N=N_1 + N_2$ is taken to be constant (the composite system is isolated), whereas both $N_1$ and $N_2$ are variables (the two subsystems can exchange particles). The entropies of both systems, 1 and 2, are now determined in the same derivation.

Swendsen starts from the probability of having $N_1$ particles in subsystem 1 and $N_2 = N - N_1$ particles in subsystem 2, which for a system of distinguishable individual particles is given by the binomial distribution
\begin{equation}\label{binom}
    P(N_1,N_2)= \frac{N!}{N_1!N_2!}(\frac{V_1}{V})^{N_{1}}(\frac{V_2}{V})^{N_{2}}.
\end{equation}
The entropy of the composite system is subsequently taken to be the logarithm of this probability, plus an arbitrary constant (that only changes the zero of the entropy scale):
\begin{equation}\label{entrop}
    S(N_1,V_1,N_2,V_2) = k \ln \frac{{V_{1}}^{N_{1}}}{N_1!} + k \ln \frac{{V_{2}}^{N_{2}}}{N_2!}.
\end{equation}
In Eq.\ (\ref{entrop}) the value of the additive constant has been set to $k \ln V^N/N!$, for reasons of convenience.  It is now clear from Eq.\ (\ref{entrop}) that the entropy of the composite system is the sum of two quantities each of which pertains to only one of the two subsystems. This suggests introducing the function
\begin{equation}\label{reducedentrop}
    S (N,V) = k \ln \frac{V^N}{N!}
\end{equation}
as a general expression for the entropy of a system of volume $V$ and particle number $N$. In the limiting situation in which Stirling's approximation for the factorials applies, taking into account that in thermodynamical equilibrium we will have $V_1/N_1= V_2/N_2$ (this corresponds to the maximum of the probability distribution), we find that
\begin{equation}\label{extensive}
   k \ln \frac{{V_{1}}^{N_{1}}}{N_1!} + k \ln \frac{{V_{2}}^{N_{2}}}{N_2!}\simeq k \ln \frac{V^N}{N!}.
\end{equation}
This leads to a nicely consistent scheme: If we were to apply the just sketched procedure for finding the entropy to the composite system itself, by combining it with a third system, we would find $S (N,V) = k \ln V^N/N!$ for the entropy of the combined system 1+2. As we now see, this entropy is equal to our earlier defined value in Eq.\ (\ref{entrop}) (fixed by adding the freely chosen constant $k \ln V^N/N!$ to the logarithm of the probability). So we obtain a consistent set of extensive entropies by taking Eq.\ (\ref{reducedentrop}) as our defining equation for entropy.

Swendsen claims that in this way the factor $1/N!$ in the formula for the entropy has been demonstrated to be a necessary consequence of the distinguishability of the gas atoms or molecules. He rejects the formula $S = k \ln W $ and maintains that Boltzmann's ideas, when pursued rigorously like in the just described argument, automatically lead to the expression $S = k \ln W/N!$.

This derivation of $S = k \ln W/N!$ is not convincing, however. First, it should be observed that its starting point, taking the entropy as $k$ times the logarithm of the probability in Eq.\ (\ref{binom}), is not really different from using the standard formula $S = k\ln W $. This is because the probability $P(N_1,N_2)$ is equal to the volume in phase space measuring the number of states with particle numbers $N_1$ and $N_2$, divided by the (constant) total number of states. So the logarithm of the probability is, apart from an additive constant, equal to the logarithm of the number of states with $N_1$ and $N_2$. Now, for the comparison with thermodynamics it suffices to replace this number of states with the total number of states: in the thermodynamic limit the probability is peaked, to an extreme degree, around the equilibrium value and the number of equilibrium states is for all practical purposes equal to the total number of states---this is explicitly used by Swendsen in his argument (e.g., Swendsen 2012). Therefore, the entropy of the composite system \`{a} la Swendsen is, apart from an additive constant, equal to $S = k \ln W$. Now, what Swendsen effectively does is to fix this additive constant as $1/N!$. There is no problem with this, and exactly the same can be done in the standard approach, since $N$---the total number of particles in the composite system 1+2---is a constant. The $N$-dependency of $S$ that is introduced here is introduced by convention, by choosing a different constant in the definition of $S$ for different values of $N$.

The next step taken in Swendsen's derivation is to require that the entropy of the system 1+2 should have the same value, and the same $N$-dependency, in the situation in which it is isolated and the situation in which $N$ is a variable (when 1+2 is brought into contact with a system 3)---this is presented as a requirement of consistency. However, this consistency requirement is exactly the condition that the entropy formula should be such that there will be no change in entropy when a partition is removed. So the derivation boils down to showing that by introducing a $N$-dependent zero in the definition of the entropy, by convention, the entropy of mixing can be eliminated. But this is what we knew all along! We were asking for a fundamental microscopic justification of the division by $N!$, but Swendsen's argument on close inspection only tells us that the division by $N!$ leads to a convenient expression that makes the entropy extensive and avoids the Gibbs paradox. The insertion of $1/N!$ is in this case just a convention.

This verdict should not be taken as a denial of the fact that the distinguishability of particles is responsible for the occurrence of factorials in expressions in which particle numbers are variables, like (\ref{binom}) and (\ref{entrop}). These factorials are important in statistical mechanics, for example in predicting what happens in mixing processes. But it was already argued by Ehrenfest and Trkal (1920, 1921; see also van Kampen 1984) that these factorials can be understood within the standard formalism and do not require a change in the formula $S = k \ln W$ for closed systems. Indeed, the dependence of the total entropy in Eq.\ (\ref{entrop}) on $N_1$ and $N_2$ is unrelated to how $N$ occurs in this formula (and to the choice of the zero of the total entropy).

\sectioneins{The difference between the thermodynamic and statistical entropies}
Our original problem was the difference in behavior between the thermodynamic and the statistical entropies: upon removal of a partition between two
containers the entropy increases according to statistical
mechanics, whereas it remains the same in thermodynamics. From the point of view of statistical mechanics there is really a change, in
the sense that the number of accessible microstates $W$ objectively
increases. \emph{In principle} we could verify this empirically, by following the paths of individual particles;
we could in this way even measure the microscopic entropy of mixing in a laboratory (Dieks 2010). Admittedly, this would require
measurements that lead us outside
the domain of thermodynamics. But from the statistical mechanics point of view these changes in phase volume and entropy must be deemed completely natural and objective. This already shows that attempts at eliminating these changes on the basis of arguments on the microscopic scale are doomed to failure. Our analysis of two of such attempts in the previous sections has confirmed this.

This leaves us with the discrepancy between the thermodynamic and statistical entropy. But is there really a problem here? Only if we think of entropy as a Platonic concept that should be the same in all cases (compare van Kampen 1984). If we accept that the two entropies are different, the problem evaporates. After all, entropy is defined
differently in statistical mechanics than in thermodynamics: in
statistical mechanics the fine-grained micro-description is
taken into account as a matter of principle, whereas in
thermodynamics this same micro-description is excluded from the
start. This difference between
the statistical mechanical and the thermodynamical approaches by itself already makes it understandable that the values of entropy changes according to statistical mechanics may sometimes be different
from those in thermodynamics (see for a discussion of the consequences of this for the second law of thermodynamics: Versteegh 2011).

From a \emph{pragmatic}
 point of view it is useful, in many circumstances, if the two
theories give us the same entropy values. We can achieve this by a
``trick'', namely by introducing a new entropy definition in
statistical mechanics: Replace $S = k \ln W$ by $S = k \ln
(W/N!)$. For systems in which $N$ is constant this makes no
difference for any empirical predictions: it only adds a
constant (though $N$-dependent!) number to the entropy value. For situations in which $N$ is made to change, this new definition leads to the disappearance of the entropy of mixing and extensivity of the statistical entropy.
In this way we obtain agreement with thermodynamics. But it is
important to realize that this ``reduced entropy'' (as it is
called by Cheng 2009) has no microscopic
foundation; rather, it may be interpreted as the result of a pragmatic decision to erase microscopic distinctions because we are not interested in them in thermodynamics. The division by $N!$ is therefore a
convention, motivated by the desire to reproduce
thermodynamical results, even though the conceptual
framework of thermodynamics is basically different from that of
statistical mechanics. The occurrence of $1/N!$ does not necessarily flow from the nature of basic properties of particles, and attempts to prove otherwise are based on a misconception. (Nor should we think that quantum mechanics makes an essential difference here: identical quantum particles can behave just as classical particles in certain circumstances, which again gives rise to the Gibbs paradox; see Dieks and Lubberdink 2011, Versteegh 2011.)

So the solution to our problem is simply to admit that there is a difference between the thermodynamic and the statistical entropy: the thermodynamic entropy is extensive, the statistical entropy is not. Given the different pictures of physical processes painted by thermodynamics and statistical mechanics, respectively, this difference is only natural.

\sectioneins{References}

%we strongly recommend to set the bibliography in the following list environment:
%citations: (Bell 1987, 23) etc.

\begin{litverzeichnis}
\item A. Ben-Naim, ``On the so-called Gibbs paradox, and on the real paradox'', in: \emph{Entropy} \textbf{9}, pp.\ 132-136 2007.
\item L. Boltzmann, ``\"{U}ber die Beziehung zwischen dem zweiten Hauptsatze der mechanischen W\"{a}rmetheorie und der Wahrscheinlichkeitsrechnung resp. den S\"{a}tzen \"{u}ber das W\"{a}rmegleichgewicht'', in: \emph{Wissenschaftliche Abhandlungen}, Volume II, pp.\ 164-224. Providence: AMS Chelsea Publishing 2001.
\item C.-H. Cheng, ``Thermodynamics of the system of distinguishable particles'', in: \emph{Entropy} \textbf{11}, pp.\ 326-333  2009.
\item D. Dieks, ``The Gibbs Paradox Revisited'', in: \emph{Explanation, Prediction and Confirmation}, edited by D. Dieks et al., pp.\ 367-377. New York: Springer 2010.
\item D. Dieks and A. Lubberdink, ``How Classical Particles Emerge from the Quantum World'', in: \emph{Foundations of Physics} \textbf{41}, pp.\ 1041-1064 2011.
\item P. Ehrenfest and V. Trkal, ``Afleiding van het dissociatie-evenwicht uit de theorie der quanta en een daarop gebaseerde berekening van de chemische constanten'', in: \emph{Verslagen der Koninklijke Akademie van Wetenschappen, Amsterdam} \textbf{28}, pp.\ 906-929 1920; ``Ableitung des Dissoziationsgleichgewichtes aus der Quantentheorie und darauf beruhende Berechnung der chemischen Konstanten'' \emph{Annalen der Physik} \textbf{65}, pp.\ 609-628 1921.
\item S. Fujita, ``On the indistinguishability of Classical Particles'', in: \emph{Foundations of Physics} \textbf{21}, pp.\ 439-457 1991.
\item D. Hestenes, ``Entropy and Indistinguishability'', in: \emph{American Journal of Physics} 38, pp.\ 840-845 1970.
\item K. Huang, \textit{Statistical Mechanics}. New York: Wiley 1963.
\item J.F. Nagle, ``Regarding the Entropy of Distinguishable Particles'', in: \emph{Journal of Statistical Physics} 117, pp.\ 1047-1062 2004.
\item S. Saunders, ``On the explanation for quantum statistics'', in: \emph{Studies in the History and Philosophy of Modern Physics} \textbf{37}, pp.\ 192-211 2006.
\item E. Schr\"{o}dinger, \textit{Statistical Thermodynamics}. Cambridge: Cambridge University Press 1948.
\item D.V. Schroeder, \textit{An Introduction to Thermal Physics}. San Francisco: Addison Wesley Longman 2000.
\item A. Sommerfeld, \textit{Thermodynamik und Statistik}. Thun: Deutsch 1977.
\item R.H. Swendsen, ``Statistical Mechanics of Classical Systems with Distinguishable Particles'', in: \emph{Journal of Statistical Physics} \textbf{107}, pp.\ 1143-1166 2002.
\item R.H. Swendsen, ``Gibbs' Paradox and the Definition of Entropy'', in: \emph{Entropy} \textbf{10}, pp.\ 15-18 2008.
\item R.H. Swendsen, ``Choosing a Definition of Entropy that Works'', in: \emph{Foundations of Physics} \textbf{42}, pp.\ 582-593 2012.
\item N.G. van Kampen, ``The Gibbs Paradox'', in: \textit{Essays in Theoretical Physics}, edited by W.E. Parry, pp.\ 303-312. Oxford: Pergamon Press 1984.
\item M.A.M. Versteegh and D. Dieks, ``The Gibbs Paradox and the Distinguishability of Identical Particles'', in: \emph{American Journal of Physics} \textbf{79}, pp.\ 741-746 2011.
\item G.H. Wannier, \textit{Statistical Physics}. New York: Wiley 1966.

\end{litverzeichnis}

\end{document}